\documentclass[12pt]{article}
\usepackage{amsfonts}

\renewcommand{\baselinestretch}{1.30}

\newcommand{\half}{\frac{1}{2}}

\def\a{\alpha}
\def\l{\lambda}

\def\b{\beta}

\def\g{\gamma}
\def\G{\Gamma}
\def\d{\delta}

\def\e{\epsilon}

\def\o{\omega}

\def\t{\theta}

\def\P{\Pi}

\def\p{\partial}
\def\pb{\bar{\partial}}

\def\be{\begin{equation}}
\def\ee{\end{equation}}
\def\ba{\begin{array}} \def\ea{\end{array}}
\def\bea{\begin{eqnarray}}
\def\eea{\end{eqnarray}}

\begin{document}

\begin{flushright}
EDO-EP-51\\
DFPD 05/TH/20\\
October, 2005\\
hep-th/0510223
\end{flushright}
\vspace{20pt}

\begin{center}
\renewcommand{\baselinestretch}{1.0}

{\large\bf THE b-FIELD 
IN  PURE SPINOR QUANTIZATION OF SUPERSTRINGS 
\footnote{Talk given by M. T. at the International Workshop 
"Supersymmetries and Quantum Symmetries" (SQS'05), Dubna, 27-31 
July 2005.}}\\
\end{center}
\begin{center}
\normalsize
\vspace{5mm}

Ichiro Oda
          \footnote{
          E-mail address:\ ioda@edogawa-u.ac.jp
                  }
\\

\vspace{5mm}
          Edogawa University,
          474 Komaki, Nagareyama City, Chiba 270-0198, Japan\\

\vspace{5mm}

and

\vspace{5mm}

Mario Tonin
          \footnote{
          E-mail address:\ mario.tonin@pd.infn.it
                  }
\\
\vspace{5mm}
          Dipartimento di Fisica, Universita degli Studi di Padova,\\
          Instituto Nazionale di Fisica Nucleare, Sezione di Padova,\\
          Via F. Marzolo 8, 35131 Padova, Italy\\

\end{center}

\vspace{5mm}
\begin{abstract}
In the framework of the pure spinor approach of superstring theories, we 
describe the Y-formalism and use it to compute the picture raised b-field.
At the end we discuss briefly the new, non-minimal formalism of Berkovits and 
the related non-minimal b-field.
\end{abstract}

The new superstring formulation of Berkovits \cite{Ber1}-\cite{Ber5}, based 
on pure spinors, has solved the old problem of  
quantization of superstrings with manifest super-Poincar\'e invariance. It 
can be considered at present as a complete and consistent formulation of   
superstring theories, alternative to the NSR and GS ones that shares 
the advantages of these two formulations without suffering from their 
disadvantages.\\

To be specific, let us consider the heterotic string.
The pure spinor approach  is based on the BRST charge
 $$ Q = \oint dz (\l^\a d_\a),  \eqno{(1)} $$
and the action
 $$ I = \int d^2z ( {{1}\over{2}}\p X^a \pb X_a +p_\a\pb\t^\a - \o_\a\pb\l^\a)
 + S_{left}, \eqno{(2)}  $$ 
where the ghost  $\l^\a$  is a pure spinor satisfying an equation
 $$ (\l\G^a\l) = 0. \eqno{(3)} $$  
Moreover, $\P^a = \partial X^a + ...$ and  $d_\a = p_\a + ... $ 
 are the supersymmetrized momenta of the superspace 
coordinates  $ Z^M = (X^m, \t^\mu)$  and  $\o_\a $  is 
the momentum of $\l^\a$. Due to the pure spinor constraint, the action $I$ is 
invariant under the local $\o$-symmetry  $$ \d\o_\a =\e_a(\G^a\l)_\a .\eqno
{(4)} $$ 
Finally, $S_{left}$  is the action for the heterotic fermions.
(For type II superstrings, $S_{left}$ is the (free) action of the 
left-handed pairs $(\hat p_\a, \hat \t^\a)$ and $(\hat\o_\a, \hat\l^\a)$, and 
one must add to $Q$ the left-handed BRST charge $\hat Q = \oint
 \hat \l^\a \hat d_\a $.)\\

Taking into account the pure spinor constraint, the action $I$ describes a 
critical string with vanishing central charge  and the BRST charge $Q$ is 
nilpotent. Moreover it has been proved \cite {Ber2}-\cite{Ber3} that the 
cohomology of $Q$ reproduces the correct physical spectrum. The recipe to 
compute  tree amplitudes \cite{Ber4} and higher-loop amplitudes \cite{Ber5}
was proposed and all the checks done untill now give support to the full 
consistency of this formulation.\\

The statement that the pure spinor approach provides a super-Poincar\'e
covariant quantization of superstring theories is correct but deserves a 
warning. The non-standard pure spinor constraint, which is assumed to hold
in a strong sense $\footnote{For  different strategies, see Refs.
\cite{Grassi1}, \cite{Kazama1}, \cite {Chest}.}$  and implies that only 11 of the 16 components 
of $\l$ are independent, gives rise to the following problems:
\begin{itemize}
\item[ i)] The $\o-\l$ OPE  cannot 
be a standard free  OPE since $ \o_\a(y)(\l\G^a\l)(z) \neq 0. $  
\item[ ii)] The $\o$-symmetry requires to be gauge fixed but the  gauge fixing
 cannot be done in a covariant way. The only  gauge invariant
fields involving $\o$  are the ghost current $J$, the Lorentz current $N^{ab}$
 and the stress-energy tensor $T^{\o\l}$
 for the  $(\o,\l)$ system. At the classical level they are respectively
 $ J = (\o\l)$ ,   
 $ N^{ab} = \half (\o\G^{ab}\l)$ and $ T^{\o\l} = (\o\p\l)$. Notice  that all
 of them have ghost number zero.
\item[ iii)] In the pure spinor approach, the antighost  $b$ (ghost number $-1$),
needed to compute higher-loop amplitudes, is a compound field which cannot 
be written in a Lorentz 
invariant way. Indeed $\o$ is the only field with negative ghost number but it 
can arise only in gauge invariant compound fields with zero (or positive) 
ghost number.
\end{itemize}
  {}From i), ii) and iii) a violation of (target space) Lorentz symmetry, at 
intermediate steps, seems to be unavoidable.
Indeed in  \cite{Ber1},\cite{Ber4} the pure spinor constraint is resolved, 
thereby 
breaking SO(10) to U(5), and  a U(5) formalism is used to compute the OPE's 
between 
gauge invariant quantities. Here we would like to describe a different but 
related approach, the so called Y-formalism, that proved to be useful to 
compute OPE's and to deal with the b-field \cite{Matone},\cite{Oda2}.\\   

Let us define the \underline{non-covariant} spinor 
$$ Y_\a = {\frac{v_\a}{(v\l)}}, $$ 
where   $v_\a$  is a constant pure spinor, so that  
$$ (Y\l) = 1. $$   (and  $ (Y\G^aY) = 0 $  ). Then consider
the projector  $$ K_{\a}^{\quad\b} = {1\over{2}} (\G^a \l)_\a( Y\G_a)^\b. 
\eqno{(5)}
 $$ that projects a 5-D subspace of the 16-D spinorial space (since 
$ Tr K = 5$).
One has  $$ (\l\G^a\l) = 0 \Longleftrightarrow \l^\a K_\a^{\quad\b} =
 0, $$  (so that $\l $   has 11 independent components and 5 
components of  $\o$  are pure gauge) and  $$
(1 - K)_\a^{\quad\b}(\G^a \l)_\b = 0. $$ 
 Using this formalism, the correct $\o-\l$ OPE is  
$$ \o_{\a}(y)\l^{\b}(z) = {\frac{(1-K(z))_{\a}^{\quad\b}}{(y - z)}}. \eqno{(6)}  $$ 
(Indeed with this equation, we obtain the OPE $\o(y)(\l\G^a\l)(z) = 0 $.)\\

Using these rules (as well as free field OPE's for $X^m$ and $(p,\t)$) one can 
compute all OPE's for composite fields and in particular for the covariant and
gauge invariant fields involving $\o$ (when they are suitably defined).
Indeed, if  $Y_\a$  enters into the game,  $\p Y_\a$ 
 has the same ghost number and conformal weight as $\o$, and as a result
in the definitions of $J$, $N^{ab}$ and $T^{\o\l} $  terms like $(\p Y,\l)$, $(\p Y \G^{ab}\l)$, 
$\p(Y\G^{ab}\l) $ etc. can arise. The coefficients of these 
$Y$-dependent terms are fixed by requiring that the algebra of OPE 's 
closes, i.e., that these spurious terms do not arise in the r.h.s. of OPE 
's. With the choice  $$ N^{ab} = \half\lbrack (\o\G^{ab}\l) +\half   
(\p Y\G^{ab}\l) - 2\p(Y\G^{ab}\l)\rbrack, \eqno{(7)} $$ $$ J = (\o\l) - 
{7\over 2}(\p Y \l), \eqno{(8)} $$ $$ T^{\o\l} = (\o\p\l) + {3\over 2} 
\p(Y\p\l), \eqno{(9)} $$  one 
recovers \cite {Oda2} the correct OPE's  with the right levels ($-3$ for 
$N$, $-4$ for $J$)
and ghost anomaly $8$, as first given by Berkovits in the 
U(5)-formalism.
Notice that all the Y-dependent terms in $N^{ab}$, $J$ and $T^{\o\l}$ are BRST 
exact. In conclusion,  $J$, $N^{ab}$ and $T^{\o\l}$, defined in eqs.(7)-(9)
 are primary and Lorentz covariant fields, and their OPE's are 
the right ones with correct central charges, levels and  ghost-number anomaly.\\

Now let us come back to the b-field.
$b$ is a field with ghost number $-1$ and weight $2$ which is essential to 
compute higher-loop amplitudes. It 
satisfies the important condition   $$ \lbrace Q,b \rbrace = T,   \eqno{(10)}$$ 
 where  $T$  is the stress-energy tensor. In the pure spinor 
approach the recipe to compute higher loops \cite{Ber5}
 is based on three ingredients:
\begin{itemize}
\item[i)] A Lorentz invariant measure factor for pure spinor ghosts.
\item[ii)] BRST closed, picture changing operators (PCO) to absorb the 
zero modes of the bosonic ghosts, that is,  $Y_C$  for the $11$ zero modes 
of $\l$ and  $ Z_B , Z_J $  for the  $11g$ zero modes of $\o$ at genus $g$.
\item[iii)]  $3g-3$  insertions of the b-field folded into Beltrami parameters 
$\mu(z,\bar{z})$, i.e., $ b[\mu] = \int d^2 z  b (z)\mu(z) $ 
at genus $g > 1$ ($1$ at genus $1$ and $0$ at tree level).
\end{itemize}
At a schematic level, the recipe for computing N-point amplitudes, at genus 
$g$ ($g \ge 2 $)(for tipe II closed superstrings), is
$$ {\cal A} = \int d^{3g-3}\tau <| \prod_{i=1}^{3g-3} b[\mu_i] 
\prod_{j=1}^{10g}
Z_{B_j}(z_j) \prod_{h=1}^g Z_J(z_h) \prod_{r=1}^{11}Y_{C_r}(z_r) |^2
\prod_{s=1}^{N} \int d^2 z_s U(z_s)>,  $$
where $\tau $ are Teichmuller parameters, $\int U $ are integrated vertex 
operators and $ < \quad \quad > $ denotes the path integral measure (that 
we shall not discuss here). For $g=1$, one integrated vertex is replaced by
one unintegrated vertex V and there is only one b-insertion. At $g=0$, 
three integrated vertices are replaced by unintegrated ones. 

In standard string theories, $b$ is the antighost of diffeomorphism.
In pure spinor approach, in the absence of diff. ghosts, $b$ is a compound 
field, which, as already noted, cannot be written as a Lorentz scalar.
Using the Y-formalism, an expression for $b$ that satisfies the fundamental 
condition (10), is \cite{Ber4}
$$ b =\half (Y\P^a\G_a 
d) + (\tilde\o\p \t) \equiv Y_\a G^\a, \eqno{(11)}   $$
where $\tilde\o$ is 
the non-covariant but gauge-invariant ghost  
$$ \tilde\o_\a = (1 - K)_\a^{\quad\b} \o_\b, \eqno{(12)} $$ and
 $$G^\a =
 \half :\P^a(\G_a d)^\a: - {1\over{4}} N_{ab}(\G^{ab}
\p\t)^\a -{1\over{4}}J\p\t^\a  - {1\over{4}}\p^2\t^\a, \eqno{(13)} $$ 
the last term in the r.h.s. of (13) coming from normal ordering.
Whereas  $G^\a$  is Lorentz covariant, $b$, due to its 
dependence on $Y_\a $, is not Lorentz invariant. However,
it turns out that the Lorentz variation 
of  $b$  is BRST exact.
In an attempt to understand the origin of the pure spinor approach 
\cite{Matone}
the b-field (11) has
been interpreted as the twisted current of the second w.s. susy charge of an 
N=2 superembedding approach, the first twisted 
charge being the BRST charge of the pure spinor approach. Even if this analysis
was done only at a classical level (and only for the heterotic string), it is 
suggestive of an N=2 topological origin of the pure spinor approach. 
The singularity of $b$  at $ (v\l) = 0 $ due to its dependence on $Y_\a$ is 
problematic in presence of the 
picture changing operators $Y_C = C_\a\theta^\a \d ( C_\b \l^\b) $ that cancel
the zero modes of $\l$, $C_\a$ being a constant spinor.
Therefore this b-field does not seem suitable to compute higher 
loops.\\
 
Since covariant and $\o$-invariant fields with ghost number $-1$, needed to 
get a b-field, do not exist,  the idea of 
Berkovits \cite{Ber5} was to combine 
$T$ with a picture raising operator $Z_B$ with ghost number $+1$ and use as 
insertion, a picture raised, compound field  $b_B$  such that  $$
\lbrace Q,b_B\rbrace = T Z_B. \eqno{(14)} $$  Then, this $b_B$ makes
it possible to define a bilocal field  $\tilde b_{B}(y,z)$ \cite{Ber5} 
such that $$ 
\lbrace Q,\tilde b_{B}(y,z)\rbrace = T(y) Z_{B}(z). \eqno{(15)} $$  
Then $3g - 3$ $ b[\mu]$ insertions ($1$ at $g=1$), together with $3g-3$ 
picture-raising operators $Z_B $ ($1$ at $g=1$), are replaced by $3g-3$ 
($1$ at $g=1$) insertions of the newly-introduced $\tilde b_{B}(y,z)$  
folded into Beltrami parameters.\\

To explain this recipe we need more details about 
the picture raising operators $ Z =(Z_B, Z_J)$   that absorb 
the zero modes of $\o$ included in $N^{ab}$ and $J$ : $$ Z_B  = \half 
(\l \G^{ab} d)B_{ab} \d(N^{cd}B_{cd}),  $$ $$ Z_J = (\l^\a
 d_\a)\d(J), $$ where $B_{ab}$ is an antisymmetric constant tensor.
Then in general $$Z = \l^\a Z_\a, $$  and  $$ \lbrace 
Q,Z \rbrace = 0. $$  It follows (by explicit computation or from 
general arguments plus pure spinor constraint) that: $$\lbrace Q,Z_\a
\rbrace = \l^\b Z_{\b \a},$$
$$\lbrace Q,Z_{\b \a}\rbrace = \l^\g Z_{\g \b \a}, $$
$$\lbrace Q,Z_{\g\b\a}\rbrace = \l^\d Z_{\d \g \b \a} + \p \l^\d 
\Upsilon_{
\d \g \b \a}, $$ where  $ Z_{\b \a}, Z_{\g \b \a}, Z_{\d \g \b 
\a}$ and 
$\Upsilon_{\d \g \b \a}$  are  $\G_{5}$-traceless,
i.e., they vanish when saturated with $
(\G_{a_1...a_5})^{\a_i\a_{i+1}}$ between two adjacent indices.
Their expressions can be found in \cite{Ber5} or \cite{Oda2}. \\
Moreover $\p Z_B$ and $\p Z_J$ are BRST exact.\\

As shown by Berkovits \cite{Ber5}, starting from $G^\a$
there exist  fields $H^{\a\b}, K^{\a\b\g}, L^{\a\b\g\d}$ 
(and $S^{\a\b\g}$) defined modulo  $\G_1$-traceless terms 
(that is modulo fields $ h_{i}^{\a_{1}..(\a_{i},\a_{i+1})..\a_{n}}$ 
which vanish if saturated with $\G^a_{\a_i\a_{i+1}}$), such that  
$$\lbrace Q, G^\a \rbrace = \l^\a T,\eqno{(16)} $$ 
$$ \lbrace Q, H^{\a\b} \rbrace = \l^\a G^\b + ..., \eqno{(17)}  $$ 
$$ \lbrace Q, K^{\a\b\g} \rbrace = \l^\a H^{\b\g} + ...,\eqno{(18)}  $$ 
$$ \lbrace Q, L^{\a\b\g\d} \rbrace = \l^\a K^{\b\g\d} + ..., \eqno{(19)} $$  
where the dots denote $\G_1$-traceless terms. Moreover, since we have
$\l^\a L^{\b\g\d\e} =0 + ...$, an equation $$ L^{\a\b\g\d} 
= \l^\a S^{\b\g\d} + ...,$$ is obtained.
Then the picture raised b-field that satisfies eq.(14)  is $$ b_B = b_1 + b_2
+ b_3 + b^{(a)}_4 + b^{(b)}_4, \eqno{(20)} $$ 
where $$ b_1 = G^\b Z_\b, \qquad b_2 = H^{\b\g} Z_{\b\g}, 
\qquad b_3 =- K^{\a\b\g}Z_{\a\b\g}.   $$ 
$$ b^{(a)}_4 = - L^{\a\b\g\d}Z_{\a\b\g\d}, \qquad
b^{(b)}_4 = - S^{\a\b\g}\p\l^\d \Upsilon_{\d\a\b\g}.  $$  

The expression of $b_B$ is quite complicated 
and Berkovits in \cite{Ber5} presented only the expressions of $G^\a$  and  
$ H^{\a\b}$.
 The technical device of using the non-covariant $Y_\a$  
as an intermediate step helps us to obtain the full expression of $b_B$
with a reasonable effort \cite{Oda1}, \cite{Oda2}.
In  order to compute
$ H^{\a\b}, K^{\a\b\g}, S^{\a\b\g}$   and  $L^{\a\b\g\d} $ 
one  makes the ansatz such that these fields can be constructed 
using only the building blocks  $$ \l^\a, \quad (\G_{a b}\l)^\a, \quad
(\G_{a}\tilde\o)^\a, \quad (\G_{a} d)^\a, $$  (as well as 
{$\P^a$} in $H^{\a\b}$); then one writes their most general expressions 
in terms of these blocks and imposes the condition that in the superfields 
$H$ and $K$ any dependence on $Y_\a$( which is implicit in $\tilde\o$) should 
be absent; then one requires that these superfields satisfy the recursive 
equations  (17) - (19). Consequently, we have found
$$ H^{\a\b} = - {1\over{16}} (\G^a d)^\a(\G_a d)^\b -
\half \l^\a\P^a(\G_a\tilde\o)^\b + 
{1\over{16}}\lbrack{\P^{a}(\G_{b}\G_{a}\l)^\a(\G^{b}\tilde\o)^{\b} - (\a 
\leftrightarrow \b)}\rbrack + ..., \eqno{(21)} $$
$$ K^{\a\b\g} = {1\over{16}} \l^\a(\G^a\tilde\o)^\b(\G_a d)^\g  + 
{1\over{32}}\lbrack{(\tilde\o\G^a)^{\a}\l^{\b}(\G_a d)^{\g}  + (\a 
\leftrightarrow \g)}\rbrack $$  $$ + 
{1\over{96}}\lbrack{(\tilde\o\G^{a})^{\a}(
\G_{ab}\l)^{\b}(\G_a d)^{\g} - (\a \leftrightarrow \g) }\rbrack + ..., \eqno
{(22)} $$
$$ S^{\a\b\g} = - {1\over{32}}(\tilde\o)^{\a}\l^{\b}(\G_a\tilde\o)^{\g} - 
{1\over{96}}(\tilde\o\G^a)^{\a}(\G_{ab}\l)^{\b}(\G^{b}\tilde\o)^{\g} +
..., \eqno{(23)} $$
and
 $$ L^{\a\b\g\d} = \l^{\a}S^{\b\g\d} + ..., \eqno{(24)} $$
where again the dots denote $\G_1$-traceless terms.\\

All these expressions are invariant under $\o$-symmetry (since $\tilde\o$ 
is invariant). Moreover $H$ and $K$ are Lorentz covariant (being 
independent of $Y_\a$ ) and therefore they depend on 
$\o$ only through $J$ 
and $N^{ab}$. Indeed, modulo $\G_1$-traceless terms,  the previous 
expressions of $H$ and $K$ can be rewritten as \\ 
$$ H^{\a\b} = {\frac{1}{16}}(\G_{a})^{\a\b}(N^{ab}\P_{b} 
- \half J \P_{a})+
{1 \over {384}} (\G_{abc})^{\a\b}\lbrack{(d\G^{abc}d)  + 24 
N^{ab}\P^{c}}\rbrack  + {1\over{8}}(\G_a)^{\a\b} \p \P^a, \eqno{(25)}  $$
which coincides with the result of Berkovits and \\ 
$$  K^{\alpha\beta\gamma} =  - {1\over 48} (\Gamma_a)^{\alpha\beta} 
(\Gamma_b d)^{\gamma} N^{ab}
- {1\over 192} (\Gamma_{abc})^{\alpha\beta} (\Gamma^a d)^{\gamma} N^{bc} 
$$  $$ +
{1\over {192}} (\Gamma_a)^{\gamma\beta} \Big[ (\Gamma_b d)^\alpha N^{ab}
+ {3 \over 2} (\Gamma^a d)^\alpha J \Big]
 + {1\over 192} (\Gamma_{abc})^{\gamma\beta} (\Gamma^a d)^{\alpha} 
N^{bc} - {1\over{32}}\G_{a}^{\b\g}(\G^a\p d)^\a . \eqno{(26)}  $$
Again the last terms in the r.h.s. of eqs.(25) and (26) come from normal 
ordering.\\
 
$ L^{\a\b\g\d}$  and   $S^{\a\b\g} $  have a residual 
dependence on $Y$. However,  when  $S^{\b\g\d}$  is saturated 
with  $\p\l^\e
\Upsilon_{\e\b\g\d}$  to get   $b^{(b)}_4$,  this 
dependence on  $Y$ drops out so that 
  $$ b^{(b)}_4 = - B_{ab}\d(B_{cd}N^{cd})\lbrack T^{\o\l} N^{ab}
+ {1\over 4} J\p N^{ab} - {1 \over 4}N^{ab}\p J   -\half N^{a}_{\; c} 
\p N^{bc} \rbrack. \eqno{(27)} $$  Furthermore, it turns out that
all the Y-dependent terms of  
$L^{\a\b\g\d}$  (linear and quadratic in Y) are  
$\G_1$-traceless  and therefore vanish when saturated with 
  $Z_{\a\b\g\d}$ so that also  $b^{(a)}_4 $  does 
not depend on  $Y_\a$.\\

It is interesting to notice the relation between the non-covariant b-field 
$b$, given in (11) and the picture raised b-field $b_B$.
Since $$ \lbrace Q,bZ \rbrace = TZ = \lbrace Q,b_B \rbrace, $$  
the quantity  $ bZ - b_B $  is closed. In \cite{Oda1}, it has been 
shown that this quantity is also BRST exact:  $$ b_B(z) = b(z)Z(z) +
 \lbrace Q,X(z) \rbrace, \eqno{(28)} $$ 
so that  $b_B$  and  $bZ$  are cohomologically equivalent.
Then, we also have $$ \tilde b_{B}(y,z) = b(y)Z(z) + \lbrace Q,\tilde X(y,z) 
\rbrace. $$  This result is interesting since it  can be used to show
that the insertion of \\  $b_{B}[\mu](z) \equiv \int \mu(y)\tilde b_B
(y,z)$  does not depend on the point $z$ of the insertion.
Indeed, since
 $\partial Z(z)$ is BRST exact, let say,
 $ \partial Z(z) = \lbrace Q,R(z) \rbrace $  and  $ \lbrace
Q,b(y)\rbrace = T(y)$  one has  $$  \partial b[\mu](z) = \int 
\mu(y)T(y)R(z)  +
\lbrace Q, \cdot \rbrace, \eqno{(29)} $$  and, modulo an exact term, the r.h.s. is 
the total derivative w.r.t. a Teichmuller parameter  $\tau$  and 
vanishes after integration over $\tau$. \\

Let us conclude this report by describing  briefly a  very 
interesting, new proposal of Berkovits \cite {Ber6}, the non-minimal pure 
spinor formalism, that in addition leads to the construction of a covariant 
b-field. The main idea behind this work
was to add to the fields involved in the pure spinor formalism a BRST quartet 
of fields $\bar \l_\a ,\bar \o^\a , r_\a, s^\a$ such that their BRST
variations are $ \d \bar \lambda_\a = r_\a$, $\d s^\a = 
\bar \o^\a$, 
$\d \bar \o^\a = 0$, $\d r_\a = 0$.  $\bar \l_\a$ is a 
bosonic 
pure spinor with ghost number $-1$,  $r_\a$ is a fermionic field that  satisfies
 the constraint $ (\bar\l \G^a r)= 0$ and
$ \bar \o^\a$  and $ s^\a$ are the  conjugate
momenta of $ \bar\l_\a$ and $r_\a$, respectively. The action is obtained by 
adding to the action $I$ in eq.(2), $\tilde I$ given by the BRST variation 
of the "Gauge fermion"  $  F = -\int (s \pb \bar \l) $ so 
that $$ I_{nm}= I + \tilde I  = \int d^2z ( {{1}\over{2}}\p X^a \pb X_a +
p_\a\pb\t^\a - \o_\a\pb\l^\a + s^\a \pb r_\a - \bar \o^\a \pb \bar\l_\a ) 
 + S_{left}. \eqno{(30)}  $$                      
This action is invariant under gauge 
symmetries involving $\bar \o$ and $s$, similar to the $\o$-symmetry so 
that, due 
to the constraints and these symmetries, each of the fields of the quartet 
has 11 components.  The new BRST charge 
is  $$ Q_{nm}= \int dz(\l^\a d_\a + \bar \o^\a r_\a), \eqno{(31)} $$ and 
the new
 (non-covariant) b-field corresponding to eq.(11) is $$ \tilde b = Y_\a G^\a + 
s^\a\partial
\bar \l_\a . \eqno{(32)} $$ 
 Of course the quartet does not contribute to the central 
charge and has trivial cohomology w.r.t. the (new) BRST charge.
Now let us define $$ b_{nm} = \tilde b + \lbrack Q_{nm}, W 
\rbrack, \eqno{(33)} $$ where $$ W = Y_\alpha 
 {{\bar\lambda_\beta} \over {(\bar\lambda \lambda)}} H^{[\alpha\beta]}
 + Y_\alpha {{\bar\lambda_\beta}\over {(\bar\lambda\lambda)^2}} r_\gamma 
 K^{[\gamma\beta\alpha]}- Y_\alpha {{\bar\lambda_\beta}
 \over {(\bar\lambda \lambda)^3}} r_\gamma r_\delta L^{[\delta \gamma 
 \beta\alpha]}, \eqno{(34)} $$ and $H^{[\a\b]},K^{[\a\b\g]}, L^{[\a\b\g\d]}$ 
are the fields defined in eqs.(21)-(26), antisymmetrized, e.g., 
$ H^{[\a\b]} = H^{\a\b} - H^{\b\a} $ 
etc. Then $$ b_{nm} = s^\a \partial\bar\l_\a + {{\bar\l_\a G^\a}\over 
{(\bar\l\l)}} +
{{\bar\l_\a r_\b H^{[\a\b]}}\over{(\bar\l\l)^2
}} - {{\bar\l_\a r_\b r_\g
K^{[\a\b\g]}}\over{(\bar\l
\l)^3}} - {{\bar\l_\a r_\b r_\g r_\d 
L^{[\a\b\g\d]}}
\over{(\bar\l
\l)^4}}, \eqno{(35)} $$ which is the new non-minimal, covariant b-field 
defined in eq.(3.11) of \cite{Ber6}.\\

As shown in \cite{Ber6}, this non-minimal formalism is nothing but 
a critical topological string, so topological methods can be applied to compute 
multiloop amplitudes where a suitable regularization factor replaces 
the picture-changing operators to deal with zero modes. The regulator proposed 
in \cite{Ber6} allows us to compute loop amplitudes up to  $g = 2$.

\end{document}